# Quasiperiodic AlGaAs superlattices for neuromorphic networks and nonlinear control systems

K.V.Malyshev[a)]

*Electronics and Laser Technology department, Bauman Moscow State Technical University, Moscow, 105005, Russia*

malyshev@bmstu.ru

The application of quasiperiodic AlGaAs superlattices as a nonlinear element of the FitzHugh–Nagumo neuromorphic network is proposed and theoretically investigated on the example of Fibonacci and figurate superlattices. The sequences of symbols for the figurate superlattices were produced by decomposition of the Fibonacci superlattices' symbolic sequences. A length of each segment of the decomposition was equal to the corresponding figurate number. It is shown that a nonlinear network based upon Fibonacci and figurate superlattices provides better parallel filtration of a half-tone picture then a network based upon traditional diodes which have cubic voltage-current characteristics. It was found that the figurate superlattice $F^0_{11}(1)$ as a nonlinear network's element provides the filtration error almost twice less than the conventional "cubic" diode. These advantages are explained by a wavelike shape of the decreasing part of the quasiperiodic superlattice's voltage-current characteristic, which leads to multistability of the network's cell. This multistability promises new interesting nonlinear dynamical phenomena. A variety of wavy forms of voltage-current characteristics opens up new interesting possibilities for quasiperiodic superlattices and especially for figurate superlattices in many areas - from nervous system modeling to nonlinear control systems development.

## I. INTRODUCTION

In recent years quasiperiodic superlattices (SLs) of finite length were attracted increasing attention due to their interesting and unexpected fundamental physical properties as well as their applications in nanoelectronics [1]. For example, dielectric quasiperiodic SLs are promising for nanophotonics (see, e.g., review [2]). Piezoelectric quasiperiodic SLs are promising for radiofrequency and microwave transmitters [3], and magnetic quasiperiodic SLs - for storage devices [4]. Semiconductor quasiperiodic SLs also have interesting properties. For example, AlGaAs quasiperiodic SLs are promising as an active media of a multicolor terahertz laser [5]. Now semiconductor quasiperiodic SLs based upon the AlGaAs are the most convenient for micro- and nanoelectronic devices due to a well-developed GaAs technology of layered heterostructures' growth.

AlGaAs SLs' quasiperiodicity appeared already in the first implementation of the quantum cascade laser (QCL) [6], where special layered nanostructures (injector and collector) were added inside of each SL's period to strengthen the coupling of adjacent cells. Quasiperiodic configurations of various spatial symmetry composed of such components as quantum dots (QD), are considered as promising for future nanoelectronics [7]. Such electronic configurations in the QD's cluster will be analogs of various p-d-hybridization's forms in the macromolecules of biological enzyme catalysts. Instrumental application of quasiperiodicity in the time domain also have a long history. Since the classical work [8] for the linear frequency modulation, linear and nonlinear radio signal's transformations in the frequency, time, angular and polarization fields are developing [9].

Famous characteristic features of semiconductor quasiperiodic SLs are highly dissected shape of an electronic states' spectrum and its self-similarity [1]. In contrast to spectra of periodic SLs and of traditional double-barrier resonant tunneling diodes (RTDs), spectra of quasiperiodic SL depend nonmonotonically upon external electric field. If we gradually increase the applied voltage, new resonant states of conduction electrons can appear suddenly. These are the states localized in 2-3 adjacent potential wells of conduction band bottom's profile across quasiperiodic SL layers those provide several close frequencies of transitions with an energy of about 10 meV in the multicolor terahertz laser in [5]. These weakly localized states can be useful for all applications that require separate closely spaced resonant peaks with a width of about 1-10 meV, and not continuous energy minibands having a width of about 100 meV. If some measures are taken against formation of strong electric field's domains in the semiconductor SL, then each such resonant peak in density of states can lead to waviness of the quasiperiodic SL's voltage-current characteristic (VCC). Therefore, quasiperiodic semiconductor superlattices can be useful in all electronic devices that are sensitive to shape of the nonlinear element's VCC.

Such devices include neuromorphic networks, in which every microelectronic neuron contains a nonlinear element such as a diode having the VCC with a negative slope. Behavior of the coupled neurons are often described by the FitzHugh–Nagumo model (see, e.g. [10], p.171). Some simplification reduces this model to the model of cellular nonlinear network (CNN), each cell of which contains just a diode, a capacitor and a few resistors - one resistor for each link with neighboring cells [11]. RTD is considered as a promising nonlinear element for such networks. By varying layers' thickness and composition in a double-barrier heterostructure, we can change shape of an initial part of the RTD's VCC. This shape is important for RF and microwave mixers, detectors and amplifiers (see for ex. [12]. p.12).

For neuron operation not the initial, but a falling VCC's part is especially important. The transition from the RTD to some quasiperiodic SL can make this falling part wavy, and it will change behavior of the neuron. In particular, VCC's waviness may be increased so that new small falling parts will appear on the VCC. This will lead to formation of new equilibrium states in the network's phase space.

Therefore, using of quasiperiodic superlattices as nonlinear elements in neuromorphic networks may be promising for a substantial transformation of their phase portraits. This is interesting for many areas - from a nervous system's modeling to the development of

information- measurement and control systems. Neuromorphic networks are performing well in parallel transformation of images. Therefore, it would be useful to clarify possible advantages of quasiperiodic SLs as nonlinear elements in some network for the parallel image conversion.

Nonlinear elements such as RTDs with VCC's various shapes are considered as promising for applications in digital circuits [13], as well as in analog signal processing [14] as modulators and amplifiers. For example, the efficiency of the power amplifier is significantly increased, if nonlinearly to distort the signal (by sharpening the upper part of the sine wave) before to amplify it. In the following cascades this nonlinear distortion can be removed by the power amplifier operating in the nonlinear saturation regime with occupying the horizontal part of its amplitude characteristic. Because of the large amplitude this method very effectively converts the energy of the DC power supply to the energy of the AC signal (see for ex. [14]. p.158).

## II. The construction of quasiperiodic superlattices for nonlinear networks

Basic principles of semiconductor quasiperiodic SL's building are described in [1], p.154. In this work Fibonacci SLs were taken as typical representatives of the quasiperiodic superlattices. In addition to them figurate SLs were constructed by decomposition of Fibonacci numbers in the sum of figurate numbers [16] and were investigated as promising network's nonlinear elements. Fibonacci number $S_N$ of rank $N$ is formed by adding $S_N = S_{N-1} + S_{N-2}$ the numbers of the two previous ranks $S_{N-1}$ and $S_{N-2}$ starting with $S_1 = 1$ and $S_2 = 1$. Similarly symbol sequence for $N$-th rank's Fibonacci superlattice $S_N$ (denoted by the same letter as the corresponding Fibonacci number) is formed by the serial connection (concatenation) $S_N = S_{N-1} + S_{N-2}$ of the symbolic sequences for previous two ranks' superlattices $S_{N-1}$ and $S_{N-2}$, starting with $S_1$= A and $S_2$= B. For example, SL $S_5 = S_4 + S_3$= BABBA, then SL $S_6$= BABBA + BAB= BABBABAB, etc.

Like Fibonacci numbers $S_N$, figurate numbers $F^M_L(N)$ can be calculated by recurrence formulas starting from given boundary values. $M$-th order's figurate number $F^M_L(N)$ is expressed through zero order's figurate numbers $F^0_L(N)$ and $F^0_L(N-1)$ by the formula $F^M_L(N)= F^0_L(N) +M F^0_L(N-1)$. In turn, figurate number $F^0_L(N)$ is expressed recursively $F^0_L(N)= F^0_{L-1}(N) + F^0_L(N-1)$ (Pascal's triangle), which finally leads to numbers $F^0_L(0)= 1$ и $F^0_0(N)= 1$ at boundaries of the Pascal's triangle. Therefore, to construct $M$-th order's figurate SLs $F^M_L(N)$ it is sufficient to construct zero-order's SLs $F^0_L(N)$, and then use the recursion formula $F^M_L(N)= F^0_L(N) +M F^0_L(N-1)$. Here, under the multiplication SL $F^0_L(N-1)$ by number $M$ we mean repetition of $M$ copies of this SL. To build the zero-order's SLs $F^0_L(N)$, we reduce them to already obtained Fibonacci SLs. To do this, we apply the formula (1) for expansion of a Fibonacci number $S_N$ by figurate numbers $F^0_L(n)$ (see, eg., [17]).

$$S_{N+1} = \sum_{n=0}^{[N/2]} F^0_{N-2n}(n) \qquad (1)$$

Here the expression $[N/2]$ denotes an integer part of the number $N/2$. We write down the symbolic sequence for SL $S_N$. Then we assign to each number $F^0_L(n)$ in the sum (1) a segment of this symbolic sequence such that its length is equal to the number $F^0_L(n)$. For example, according to the formula (1), $S_1= F^0_0(0)$ and $S_2= F^0_1(0)$. But SL $S_1=$ A and SL $S_2=$ B, therefore we obtain SL $F^0_0(0)=$ A and SL $F^0_1(0)=$ B. Similarly we act to obtain all remaining SLs $F^0_L(N)$. Thus from the decomposition $S_3= F^0_2(0) + F^0_0(1)$ (in numbers it looks like 2=1+1) we obtain $S_3=$ BA$= F^0_2(0) + F^0_0(1) =$B+ A. Hence SL $F^0_2(0)=$ B and SL $F^0_0(1)=$ A. Further, for example, from the decomposition $S_8= F^0_7(0) + F^0_5(1) + F^0_3(2) + F^0_1(3)$ (in numbers it looks like 21= 1+ 6+ 10+ 4) we obtain BABBABABBBABBABBABBABAB = B + ABBABA+ BBABBABABB+ ABAB. Hence SL $F^0_7(0) =$ B, SL $F^0_5(1) =$ ABBABA, SL $F^0_3(2)=$ BBABBABABB, and SL $F^0_1(3)=$ ABAB. The advantage of this method for constructing of zero-order's SLs $F^0_L(N)$ is that they inherit from Fibonacci SLs their stochastic properties.

In this work for the blocks A and B we take different layered AlGaAs heterostructures with thicknesses of a few GaAs -monolayers (ML), each by 0.565 nm (layered structure on the left in Figure 1) so that total SL length does not exceed the characteristic electron mean free path of about 100 nm.

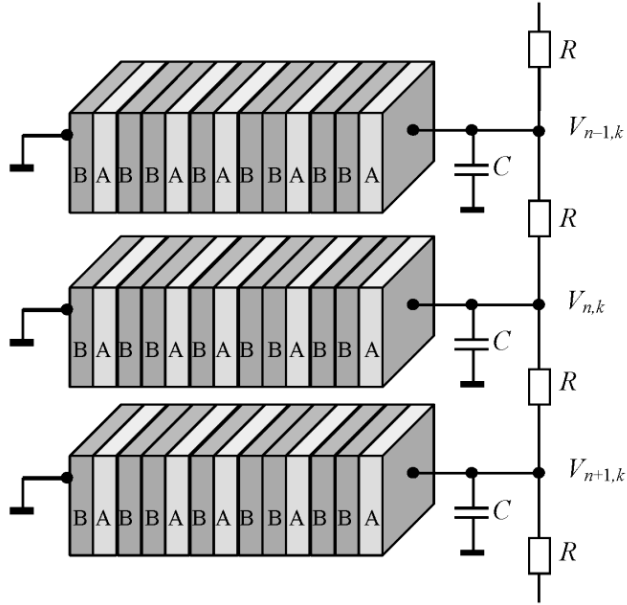

Figure.1. Three neighboring network's cells based on the Fibonacci superlattice $S_7$.

At this thickness, a probability of the appearance of strong electric field's domains is small. Therefore, states of conduction electrons are considered coherent throughout all SL's layers in a classical transfer matrix method. This assumption about the coherence of the conduction electron's wave function is not necessary to calculate a VCC using the Tsu-Esaki formula. In this formula we need to know only the electronic transparency of the whole structure, i.e. the probability of electron's transition through all barriers and wells. The same results about the transparency peaks, electron localization and resonant tunneling in the quasiperiodic semiconductor superlattices are obtained by the strong coupling method [15], that requires only knowledge of the probability of an electron jump to the next site.

Thickness and composition of this layers in blocks A and B were chosen so that SL's VCC had an elongated falling part (preferably in a wavy form) in moderate electric fields of about 10 kV/cm (Figure 2).

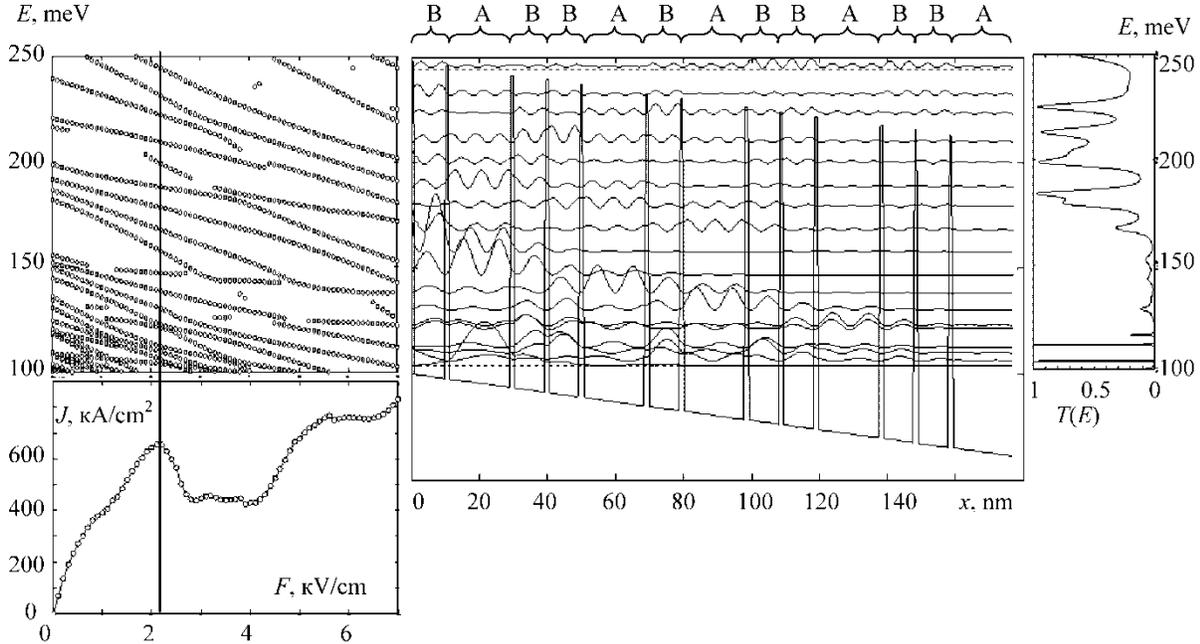

Figure 2. The characteristics of cell's nonlinear elements based upon the Fibonacci superlattice $S_7$. Bottom left – an element's voltage- current characteristic $J(F)$. The vertical line indicates the falling part's beginning. Left top - the dependence of resonance energies $E$ on the electric field strength $F$. Middle - the band diagram and squared moduli of electron's wave function depending on a coordinate $x$ across the SL's layers. Right - the dependence of the tunneling transparency $T(E)$ on the electron's energy $E$ in the electric field corresponding to the beginning of the current-voltage characteristic's falling part.

## III. Voltage-current characteristics of quasiperiodic superlattices for nonlinear networks

To calculate VCCs the formula Tsu-Esaki was used in combination with the traditional transfer matrix method (see, e.g. [18], p.58, and [19]). Correctness of computational procedures was checked by reproduction of the VCC (not shown here) one of the famous traditional double-barrier AlGaAs heterostructure [20].

Briefly, the method of calculation is as follows. For a given voltage $U$ calculation of the current density $J$ across the layers of the heterostructure is reduced to the numerical integration in the Tsu - Esaki formula (*).

$$J(U) = C \int_0^\infty dE\, T(E) F(E)$$

(*)

This formula contains two functions $F(E)$ and $T(E)$ of an electron energy $E$. The constant $C$ is determined as $C \cong em/2\pi^2\hbar^3 \approx 2 \cdot 10^9$ (A/cm$^2$)/eV$^2$ and sets the maximum theoretical current density. Here, $e$ and $m$ - charge and effective mass of the electron, $\hbar$ - Dirac's constant. The function $F(E)$ is known - it decreases slowly with increasing energy $E$ and increases monotonically with increasing voltage U according to the formula $F(E)=(1/\beta)\ln[(1+\exp(\beta(E_F-E))/(1+\exp(\beta(E_F-E-eU))]$. Here $\beta=1/k_BT_0$, where $k_B$ - Boltzmann constant, $T_0$ - temperature. The Fermi energy $E_F$ increases monotonically with increasing dopant concentration $N_d$ in the degenerate near-contact regions according to the formula $E_F =(\hbar^2/2m)(3\pi^2\,N_d)^{2/3}$. We find the transparency $T(E)$ by the numerical calculations using transfer matrix method. To do this, the $x$-axis across the heterostructure's layers we split into small parts such that within each $n$-th part the electron's potential energy $V(x)$ can be considered as constant $V_n$. Solution of the Schrödinger equation for this case is known as a sum of two plane waves $y=A_n\exp(K_nx)+ B_n\exp(-K_nx)$. Here, the wave number $K_n$ is calculated by the formula $K_n = [2m(E- V_n)]^{1/2}/\hbar$. To find the unknown coefficients $A_n$ and $B_n$ the conditions of continuity of the function $y(x)$ and of its derivative $dy/dx$ at the boundaries of adjacent small parts are used. So we obtain a system of coupled equations for neighboring $A_n$ and $B_n$. By involving the boundary conditions for the wave function far away from the heterostructure we obtain the amplitude of the wave passing across all layers. The required transparency $T(E)$ is the ratio of the transmitted flux of electrons to the incident one and is proportional to the square of the $A$'s modulus.

In this work each quasiperiodic SL's block B(A) was consisted of a barrier layer having a thickness of 2 ML, followed by a potential well with a thickness of 16(32) ML (see the band diagram in the middle of Figure 2). A height $V$ of the potential barriers (in eV) and an effective mass $M$ (in units of a free electron mass) were calculated from the traditional expressions $V = 1.11x - 0.93x^2 + 0.85x^3$ and $M = 0.067 + 0.083x$, where $x$ - aluminum's proportion in the Al$_x$Ga$_{1-x}$As- layer. This proportion was $x = 0.15$, which gave a potential barriers' height 0.15 эB for conduction electrons. At near-electrode n+GaAs- layers a Fermi energy was taken equal to 0.069 eV, and an effective mass $M = 0.067$. A contact potential difference of 0.1 eV between the n + GaAs-layers and middle undoped i-AlGaAs-layers has been added to the potential profile. A temperature was assumed equal to 300 K.

Figure 3 shows VCCs of network's nonlinear elements based on the figurate SLs $F^0_{11}(1) =$ ABBABABBABBA and $F^8_{11}(1) =$ ABBABABBABBABBBBBBBBB in comparison with the reference cubic VCC of a conventional nonlinear element like RTD. Figure 3 shows several of the same current-voltage characteristics as in the tab of the Figure 2, just shifted and scaled for the convenience of analysis of the nonlinear element's action in parallel image conversion using nonlinear network (see next section).

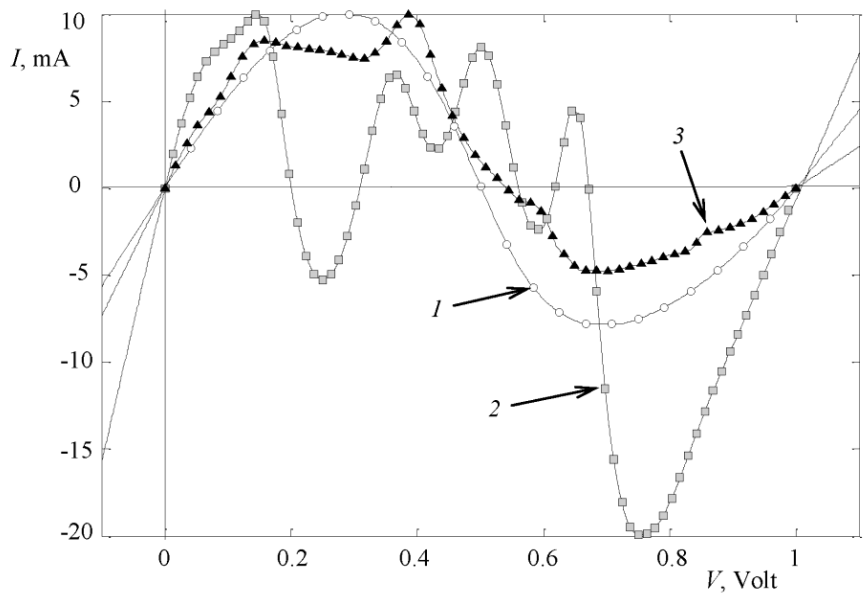

Figure 3. The voltage-current characteristics of network's nonlinear elements: *1*- reference «cubic» diode, *2*- figurate superlattice $F^0_{11}(1)$, *3*- figurate superlattice $F^8_{11}(1)$.

The reference cubic voltage-current characteristic has a symmetrical falling part with identical positive and negative regions and serves to compare of all other VCCs. VCCs are normalized so that they have the same maximum current $I = 10$ mA, and the most left (right) growing branch intersects with zero at the voltage $V = 0$ (1) Volt respectively. At a microelectronic implementation such shift of VCC's zero may be achieved by adding appropriate permanent current and voltage sources to all network's cells at once.

## IV. Parallel image transformation using nonlinear networks

We considered a nonlinear network in the form of flat lattice of the Fitzhugh-Nagumo neurons [10]. This lattice simulates some kind of a two-dimensional excitable media. A propagation of an excitation in such media is described by a system of kinetic equations, which reduces to the single reaction-diffusion equation after an assumption of the absence of a slow inhibitor [21]

$$\frac{\partial V(\vec{r},t)}{\partial t} = D_V \Delta V + F_R(V)$$

(2)

Here $V$- variable describing the excited state of the medium at space point *r* at time moment $t$, $D_V$ – diffusion coefficient describing spreading of the excitation, $\Delta$ – Laplacian, $F_R(V)$ – rate of the excitation's growth caused by an autocatalytic reaction.

To carry out a parallel signal conversion using such a medium, the signal is set as a spatial distribution $V(r,t_0)$ of the excitation at some initial moment of time $t_0$. Next, a medium evolves according to the equation (2) and after some time, we obtain the converted signal as a new spatial distribution $V(r,t)$. If desired transformation is to clean up the distribution $V(r,t_0)$ from a high-frequency spatial noise, it is doing good by the diffusion term $F_D = D_V \Delta V$ in (2). However, this term reduces not only harmful noise's amplitude, but also amplitude of useful low-frequency signal's components. This is the reaction term $F_R$ in (2) that must counteract this (Figure.4).

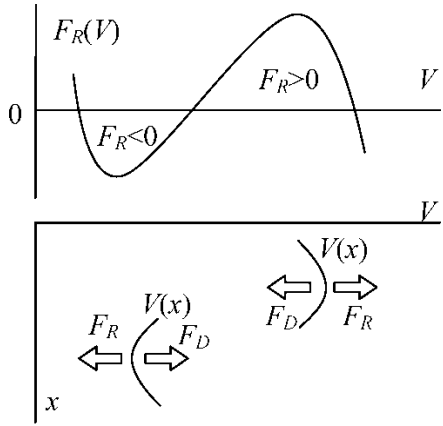

Figure.4. Opposing actions of the diffusion $F_D$ and reaction $F_R$ terms in signal $V(x)$ processing by the diffusion-reaction medium. Here $x$ - coordinate in the medium.

The diffusion term $F_D$ reduces a height of hills and a depth of wells in the relief of useful low-frequency signal's components. For preventing this phenomenon the reaction term $F_R$ should shift down wells, and shift up hills. This explains why the dependence of $F_R(V)$ must be $S$-shaped.

In a microelectronic design such two-dimensional diffusion-reaction medium can be approximately realized by means of a planar nonlinear network [22]. In nodes of this network there are cells that contain resistors, capacitors and nonlinear resistances (рис.1). Under this implementation the equation (2) takes the form (3)

$$\frac{dV_{n,k}}{dt} = \frac{V_{n-1,k} - 2V_{n,k} + V_{n+1,k}}{\tau} + \frac{V_{n,k-1} - 2V_{n,k} + V_{n,k+1}}{\tau} - \frac{I(V_{n,k})}{C}$$

(3)

Here, $V_{n,k}$ - voltage in $(n,k)$-th network's node, $\tau = RC$ - time constant of identical $RC$-chains connecting neighboring nodes. Each chain consisted of the resistance $R = 10$ Ohms and the capacitance $C = 10$ pF, which gave the characteristic time of image transformation $T = 100$ ps. In solving the system (3) at each step of time evolution the value of current $I(V)$ through a nonlinear element is calculated by cubic approximating of the VCC comprising 1,000 points. A role of the diffusion $F_D$ term of equation (2) is played by the first two terms in equation (3), and a role of the reaction $F_R$ term - by the last term in equation (3), i.e. by the nonlinear element's VCC with a minus sign. Thus, to prevent smoothing of the useful signal, the nonlinear network element's VCC must be *N*-shaped.

In a practical realization of such a network in the matrix photoconverter an image is divided into $N_x$ pixels horizontally and $N_y$ pixels vertically [23]. Under every pixel the network node is located. We assume that an initial voltage $V_{n,k}$ in the node cell (*n,k*) is equal to an intensity of light striking to this pixel, and can range from 0 (black) to 1 (white) Volt. After the initial illumination a network is disconnected from the photodetector and the voltage $V_{n,k}$ at each node cell (*n,k*) begins to vary in time *t* due to presence of a capacitance and a nonlinear element within the network's cell, and also due to the resistive connection of each cell to its neighbors. After some characteristic time, determined by the network parameters, each network's node can be connected to the corresponding node of a light-emitting matrix to obtain a transformed image. Details of image conversion using such networks are described in [22].

In this work we assumed $N_x$ = 64 points horizontally and $N_y$ = 64 points vertically. A method of network's research was used the same as in [22]. To find the dependence $V_{n,k}(t)$ we have solved a system of $N= N_x*N_y = 4056$ ordinary differential equations (3) using a specialized MatLab package's algorithm «ode15s». For boundary cells a condition of image's smoothness was assumed, i.e. a condition of continuity of a spatial derivative of the voltage in each border's cell. For example, it was assumed that $V_{0k}= 2V_{1k} - V_{2k}$ for the first row's cells in (3). Cases of indexes 0 and 1+$N_y$ ($N_x$) at the first and last rows (the first and last column) were considered similarly.

Gaussian noise in the form of a random sample of $N$ numbers having a normal distribution with zero mean and a standard deviation of 0.1 was added to the original reference image. In this work the purpose of image converting via network was the obtaining an image as close to the reference image as possible, i.e., obtaining an image purified from the added Gaussian noise. The relative standard deviation $D$ was assumed as a measure of the difference between the transformed image and the reference one. It was calculated as the ratio $|\boldsymbol{V}\text{-}\boldsymbol{W}| / |\boldsymbol{W}|$ of vectors' modules. Here $\boldsymbol{V}$ - vector of the converted image, $\boldsymbol{W}$ - vector of the reference image, $|.|$ denotes a modulus of the vector. Components of the images' vector $\boldsymbol{V}$ are the voltages $V_{nk}$ at network nodes. Images were compared at 10 consecutive time moments in the interval 0 - 2 in units of $\tau = RC$. Usually in this time interval of the order of 100 ps the conversion error $D$ first decreases and then increases for the studied VCC's shapes.

## V. Action of nonlinear networks based on quasiperiodic superlattices

The existence of many maxima in the quasiperiodic SL's VCC can lead to a presence of several stable equilibrium states of the network's cell. This multistability is important for all processes occurring in nonlinear networks. In particular, benefits of VCC with many maxima for parallel image processing using the cellular nonlinear network are described in [23]. There VCC with many maxima was obtained using several operational amplifiers and analog multipliers. Here we see that a VCC with several well distinguishable maxima can be obtained using only one quasiperiodic SL. This opens the possibility of parallel image processing in time scale of the order 10 ps.

Figure 5 shows the process of a noisy image's transformation (Fig.5*b*) using typical nonlinear network based on a reference diode with the cubic VCC shown in Figure 3.

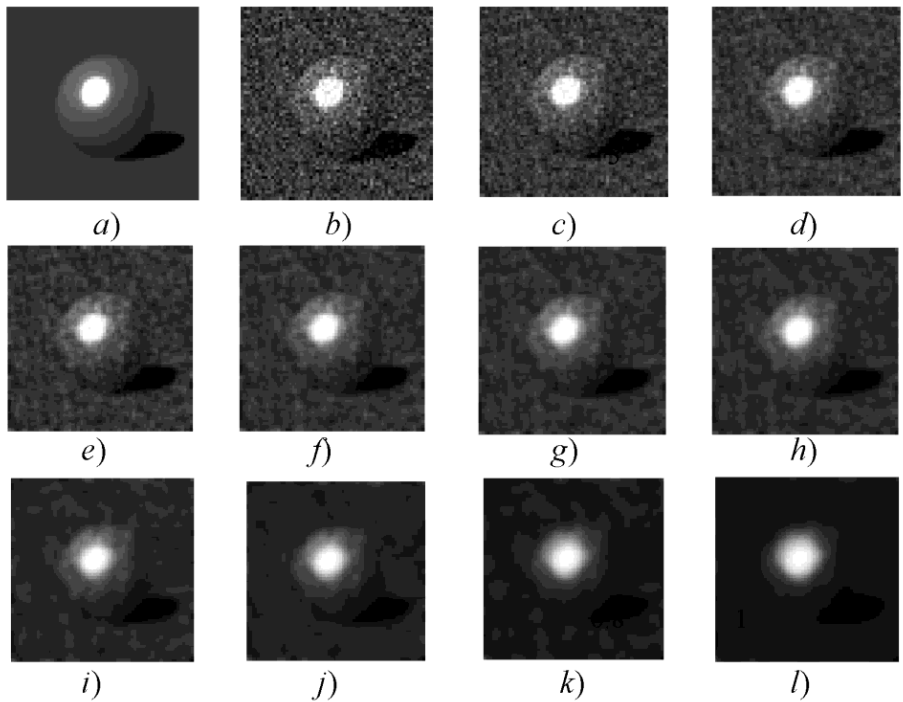

a)    b)    c)    d)

e)    f)    g)    h)

i)    j)    k)    l)

Figure 5. Converting an image using network based upon a reference diode having the cubic VCC: a) - reference image, b) - noisy image before conversion, c)…l) – converted image at the times t/τ=0.1(c), 0.2(d), 0.3(e), 0.4(f), 0.5(g), 0.6(h), 0.8(i), 1.1(j), 1.5(k), 2.0(l).

Deviation from the reference image (fig.5*a*) initially decreases (fig.5*b-d*) because of noise's smoothing, reaches a minimum (fig.5*e*) at the time point $t/\tau = 0.3$ and then increases (fig.5*b-d*) due to contrast increasing.

Figure 6 compares time dependences of the mean-square-deviation $D$ for images obtained using network based on the "cubic" diode, figurate superlattice $F^0{}_{11}(1)$ and figurate superlattice $F^8{}_{11}(1)$.

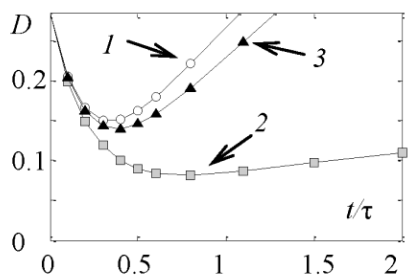

Figure 6. Time dependences of the image's deviation from the reference image during conversion using network based on the 1- diode having a cubic VCC, 2- figurate superlattice $F^0_{11}(1)$, 3- figurate superlattice $F^8_{11}(1)$.

For all structures the smallest deviations were achieved within a time comparable with a characteristic time τ. For SL $F^0_{11}(1)$ the smallest deviation of transformed image from the reference one is almost twice less than for the other structures.

On Figure 7 converted images having the smallest deviations from the reference image are compared with each other.

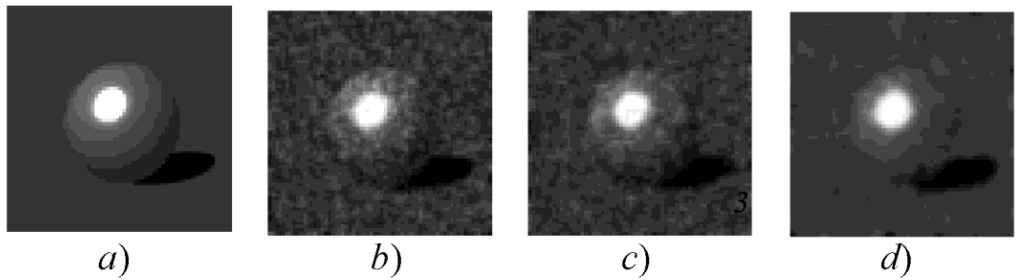

a)                b)                c)                d)

Figure 7. Converted images having the smallest deviations from the reference one: *a*) - reference image, *b*)…*d*) – best images, obtained using the network based on the: *b*) – diode having a cubic VCC, *c*) – figurate superlattice $F^8_{11}(1)$, *d*) – figurate superlattice $F^0_{11}(1)$.

It can be seen that an improvement of image transformation in this series (Fig.7b-d) is associated with the improvement of transformation of an image's background. Network based on the reference "qubic" diode leaves background grainy (Fig.7b). Network based on the figurate SL $F^8_{11}(1)$ significantly reduces a granularity of the background (Fig.7c). At last, network based on the figurate SL $F^0_{11}(1)$ almost entirely removes the background's granularity (Fig.7d).

The falling part of a nonlinear element's VCC in network's cell is responsible for the transformation of an image background. Cubic VCC of the conventional nonlinear element has one positive and one negative part in the unit voltage range (Figure 3). During transformation the VCC's positive part shifts a dark gray point of the image toward black and the VCC's negative part shifts light gray towards white. Background's points having a gray color turn out in an unstable state during this process. To improve transformation it is desirable that gray points were moving to extreme states of black and white very slowly, or that they were moving to some steady states which have approximately the same gray color. Middle parts of the VCCs of nonlinear elements based on figurate SLs just satisfy these requirements (Figure 3). Negative part of the VCC of the figurate SL $F^8_{11}(1)$ is wavy, and the positive one is non-monotonic. Therefore, the conversion of a gray background using SL $F^8_{11}(1)$ is better than using a reference "qubic" diode. Nonmonotonicity of the VCC of the SL $F^0_{11}(1)$ is so large that several positive and negative VCC's branches appears. Rising portions of these branches are crossed with horizontal at points that are stable states of gray in an image. Therefore, the conversion of a gray background using SL $F^0_{11}(1)$ turns out better than using SL $F^8_{11}(1)$ and much better than using a reference "qubic" diode.

Other quasiperiodic SLs as a part of a nonlinear network's element behave analogously to considered figurate SLs during image filtration. Image transformation depends on the waviness of a VCC's falling part and on the amplitude of this waviness. VCC of the Fibonacci SL $S_7$= BABBABABBABBA (fig.2) и VCC of the figurate SL $F^1_1(6)$= ABBABABBBABAB (not shown) have weak waviness of the falling part, so behave analogously to the figurate SL $F^8_{11}(1)$. The waviness of the VCC of the Fibonacci SL $S_8$= BABBABABBABBABABBABAB (not shown) increases but does not achieve the waviness of the VCC of the SL $F^0_{11}(1)$. So the SL $S_8$ in sense of image filtering error turns out between the SL $F^8_{11}(1)$ and the SL $F^0_{11}(1)$.

It should be noted that advantages of quasiperiodic SL associated with multistability of cells, with a high probability will manifest itself not only in neuromorphic networks, but also in all other nonlinear dynamical systems. This follows from equivalence between above discussed nonlinear network's cell and an overdamped oscillator located in a potential, the shape of which depends on the shape of the nonlinear element's VCC [23]. So the transition from conventional heterostructures having $N$-shaped VCC like RTD to quasiperiodic SLs is equivalent to transition from a dynamical system within a double-well potential to the system within a multiwell potential. Therefore, in the field of nonlinear systems, use of quasiperiodic SLs may lead to discovery of new interesting phenomena. For example, it will be interesting to investigate propagation of solitons and spiral waves in the reaction-diffusion media based on quasiperiodic SLs, as well as the phenomena of self-organization, including a self-organized criticality and chaotic oscillations. The possibly easiest way for the practical realization of the benefits of the discussed superlattice diodes - it is just to insert them in a classical digital [13] and analog [14] schemes instead of standard TD's and RTD's.

## VI. Conclusion

On the example of Fibonacci and figurate AlGaAs superlattices it is shown that quasiperiodic semiconductor superlattices are promising as nonlinear elements in cells of the FitzHugh–Nagumo neuromorphic networks. A waviness of a falling branch of a current-voltage characteristic of the quasiperiodic superlattice may lead to a new equilibrium states in the network's phase space. Arising in this way multistability has a positive effect on a process of parallel conversion of an analog signal using the neuromorphic network. In particular, figurate superlattice $F^0_{11}(1)$ as a part of a nonlinear network's element provides the conversion error almost twice less than do traditional diodes with a cubic current-voltage characteristic. Therefore, quasiperiodic superlattices are promising for nonlinear information-measuring and control systems, as well as for modeling of the nervous system. The multistability of a network's cell based on the quasiperiodic semiconductor superlattices promises new interesting nonlinear dynamical phenomena.